\documentclass[aps,prb,superscriptaddress,twocolumn]{revtex4-1}

\usepackage{color}

\usepackage{amsmath,amsfonts,amsthm}
\usepackage{chngcntr}
\usepackage{gensymb}
\usepackage{amssymb}

\usepackage[T1]{fontenc} 
\usepackage[english]{babel} 
\usepackage[ruled,noend, algo2e]{algorithm2e} 
\usepackage{multirow}
\usepackage{graphicx} 

\usepackage[]{algorithm2e}
\usepackage{lmodern}
\usepackage{mathtools}
\usepackage{hyperref}   

\usepackage{lipsum} 
\usepackage{algorithm}
\usepackage{algorithmic}
\usepackage{fancyhdr} 
\numberwithin{equation}{section} 
\numberwithin{figure}{section} 
\numberwithin{table}{section} 

\renewcommand{\thetable}{\arabic{table}}
\counterwithout{figure}{section}
\counterwithout{table}{section}

\begin{document}
\title{Accelerating computational modeling and design of high-entropy alloys} 
  
\author{Rahul Singh}
\affiliation{Department of Mechanical Engineering, Iowa State University, Ames, Iowa 50011 USA}
\author{Aayush Sharma}
\affiliation{Department of Mechanical Engineering, Iowa State University, Ames, Iowa 50011 USA}
\author{Prashant Singh}
\affiliation{Ames Laboratory, U.S. Department of Energy, Iowa State University, Ames, Iowa 50011 USA}
\author{Ganesh~Balasubramanian}
\affiliation{Department of Mechanical Engineering \& Mechanics, Lehigh University, Bethlehem, PA 18015 USA}
\author{Duane D. Johnson}
\affiliation{Ames Laboratory, U.S. Department of Energy, Iowa State University, Ames, Iowa 50011 USA}
\affiliation{Department of Materials Science \& Engineering, Iowa State University, Ames, Iowa 50011 USA}

\begin{abstract} 
{\bf With huge design spaces for unique chemical and mechanical properties, we remove a roadblock to computational design of {high-entropy alloys} using a metaheuristic hybrid Cuckoo Search (CS) for ``on-the-fly'' construction of Super-Cell Random APproximates (SCRAPs) having targeted atomic site and pair probabilities on arbitrary crystal lattices. Our hybrid-CS schema overcomes large, discrete combinatorial optimization by ultrafast global solutions that scale linearly in system size and strongly in parallel, e.g. a 4-element, 128-atom model [a $10^{73+}$ space]  is found in seconds -- a  reduction of 13,000+ over current strategies.   With model-generation eliminated as a bottleneck, computational alloy design can be performed that is currently impossible or impractical. We showcase the method for real alloys with varying short-range order. Being problem-agnostic, our hybrid-CS schema offers numerous applications in diverse fields.}
\end{abstract} 
\maketitle

\section{Introduction}
Complex solid-solution alloys (CSAs), a subset of which are near-equiatomic high-entropy alloys,\cite{Yeh2004,CB2004,Senkov2015,Singh2015,Gao2016,Singh2018,EPG2019}  show remarkable properties for number of elements $N \ge 4$ and set of elemental compositions $\{c_{\alpha={1,N}}\}$,\cite{Miracle2017} even for medium-entropy ($N$=3). These findings have encouraged research into CSAs for applications in aerospace and defense, like adding refractory elements for higher operational temperatures.  
For higher-melting refractory CSAs,  vacancy defects -- ubiquitous when processing -- can have a profound influence on stability and phase selection \cite{Singh2020}, adding another design ``element''.
Thus, CSAs have a vast design space to create materials with novel or improved properties (e.g., structural strength or resistance to fatigue, oxidation, corrosion, and wear), while other properties (e.g., resistivity, thermoelectricity, elasticity, and yield strength) can alter rapidly with a set of  $\{c_{\alpha}\}$.\cite{Singh2015,Miracle2017,Singh2018,Singh2020,Zhang2015,Karati2019,QD2019,Li2016,Zhang2020,Singh2018b}     As such, accurate ``on-the-fly'' constructed CSA models are needed to enable computational design and to identify trends in $\{c_{\alpha}\}$-derived properties and thermal stability.   Yet, models of CSAs have a design space that grows exponentially -- a NP-hard combinatorial problem. 

{\par} To remove this design roadblock, we establish a \emph{hybrid Cuckoo Search} (CS), an evolutionary algorithm\cite{TB1996} inspired by Yang and Deb\cite{Yang2009} based on the brood parasitism of a female Cuckoo bird (mimicking color and pattern of a few host species).  Importantly, the CS advantages are: (a) guaranteed global convergence, (b) local and global searches controlled by a switching parameter, and (c) L$\'{e}$vy flights scan solution space more efficiently -- no random walks, so better than a Gaussian process.\cite{Yang2009,Yang2010,Yang2013}  A CS yields approximate solutions (``nests'') for intractable or gradient-free problems\cite{Blum2003} with little problem-specific knowledge -- often only a solution ``fitness'' function.\cite{Ashlock2006} For complex cases, fitness can be discontinuous, non-differentiable to noisy.  Related methods\cite{Yang2010} are genetic-algorithm,\cite{Holland1992} simulated-annealing,\cite{Kirkpatrick1983} particle-swarm,\cite{Kennedy1995} and ant-colony\cite{Dorigo1996} optimization. 

{\par}Inspired by successes, \cite{Yang2010,Yang2011a,Yang2013} including materials design,\cite{SharmaCS2017}  our hybrid-CS schema is more efficient than CS, as we establish for standard test functions, where CS already bests all other evolutionary algorithms. Our hybrid CS employs L$\'{e}$vy flights for global optimization and Monte Carlo (MC) for local explorations of large multi-modal design space. Selecting a best nest each iteration (or cycle) ensures that solutions ultimately converge to optimality, while diversification via randomization avoids stagnation, i.e., trapped in local minima.

{\par}As CSAs have properties that can alter rapidly with composition $\{c_{\alpha}\}$, a hybrid CS enables ``on-the-fly'' optimal-model generation with exceptional reduction in solution times, scaling linearly with system size and strongly for parallel solutions (i.e., doubling processors, halves compute time). 
We construct a pseudo-optimal (discrete) Super-Cell Random APproximates (SCRAPs) with specified one- and two-site probabilities, as directly qualify a model's fitness and as can be measured.\cite{Cowley1950,Moss1964}   
For a 4-element, 128-atom model [$10^{73+}$ space], a 13,000+ reduction in execution time (optimal in 47 secs) is found over current strategies. And, only minutes are needed for a 5-element, 500-atom SCRAP [$10^{415+}$ space]!  

\begin{figure}[t]
\centering
\includegraphics[scale=0.15]{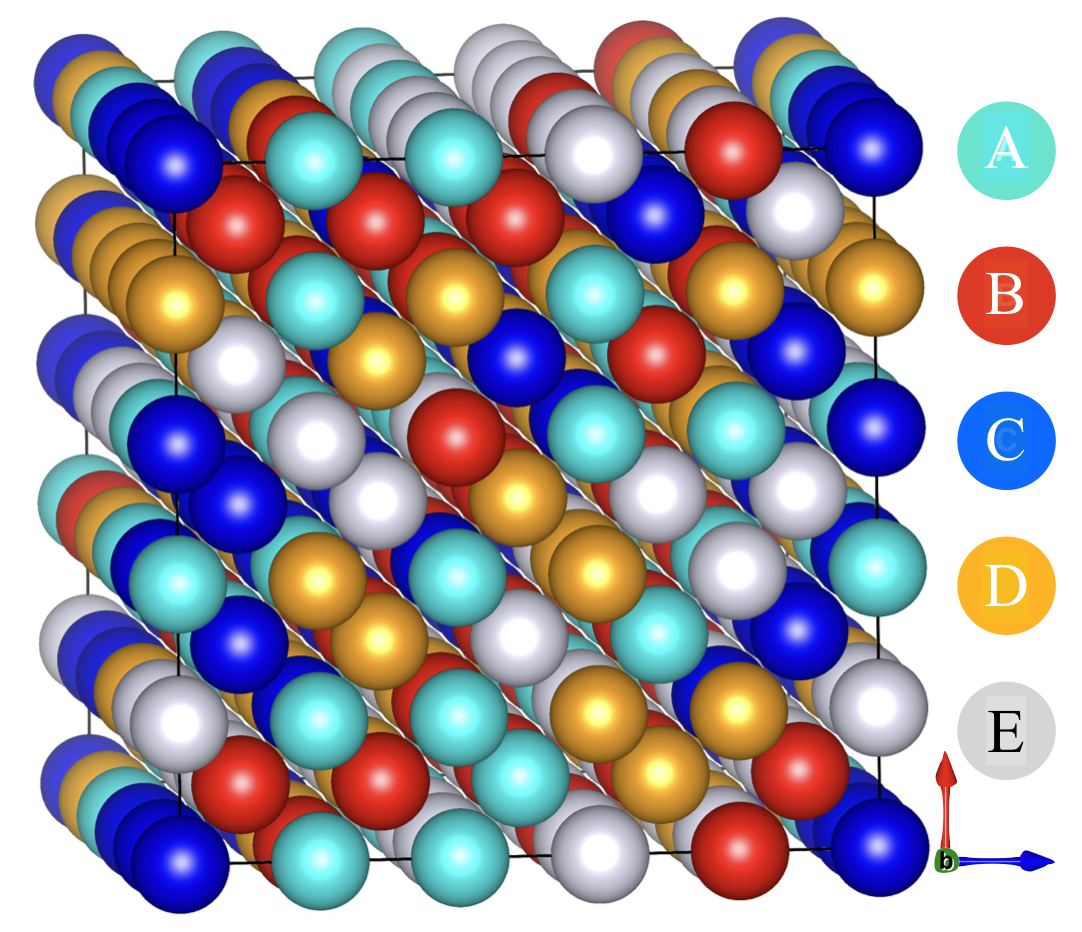}
\caption{Hybrid-CS-optimized $250$-atom cell for a bcc equiatomic ABCDE solid-solution alloy with zero SRO for 3 neighbor shells around each atom.
See Table~\ref{tab2} for timings.}
\label{fig:SCRAP5}
\end{figure}

{\par}Hybrid-CS SCRAPs are optimal for $S$ sites occupied by $N$ elements to mimic CSAs (Fig.~\ref{fig:SCRAP5}) with target one-site  $\{c_{\alpha}\}$ and two-site probabilities for a crystal symmetry, e.g., body- (bcc) or face- (fcc) centered  cubic. Two-site probabilities are atomic short-range order (SRO) parameters that qualify a model's  fitness, with values over range $R$ (say, $1$-$5$ neighbor shells) with $Z_R$ atoms per shell, for a total per site of {\small$\frac{1}{2}N(N-1) \cdot \sum_R Z_R$}, as can be measured experimentally.\cite{Cowley1950,Moss1964} 
Solution spaces grow exponentially:  a 5-element equiatomic alloy in a 250 (500)-atom cell has an estimated space of $10^{169}$ ($10^{415}$). 
Hybrid CS optimizes for very large problems with extraordinary reduction in times over current methods (Table~\ref{tab2}),  eliminates the model-generation bottleneck, and permit computational design that is currently impossible or impractical.

{\par}After establishing the \emph{bona fides} of  hybrid CS, we define fitness and associated physical (and discrete bounds) to eliminate stagnation of  MC searches.  Hybrid-CS-generated SCRAPs are presented for CSAs with targeted SRO in different crystal structures to prove solution times scale linearly with size and strongly for parallel solution and to show the hybrid~CS enables rapid design of optimal ``nests'' (SCRAPs).   We then showcase design and model assessment using density-functional theory (DFT) to predict properties. For any random alloy, we discuss a symmetry requirement that permits a reduction in the number of DFT calculations for design. Notably, however, the hybrid CS is problem-agnostic, and offers far-reaching applications to problems in manufacturing, finance, commerce,  science, and engineering. 

\section{Results}
\subsection{Hybrid CS vs CS}
Hybrid CS reaps benefits of MC for local optimization alongside CS for global optimum using multiple-nest explorations via L$\'{e}$vy flight.   A ``nest'' represents, for example, a function value or an alloy configuration (SCRAP). A global CS discards a fraction of nests, $p_a$, with worst fitness (i.e.,  probability of finding an alien nest\cite{Yang2009}).   As shown in the pseudo-codes in Methods, we replace the local search in CS Algorithm~\ref{Algorithm1} by MC and create a hybrid CS Algorithm~\ref{Algorithm2}. We tested various MC approaches, including simulated annealing, and the basic version, as embodied in Algorithm~\ref{Algorithm2}, was superior to all others.

\begin{figure}[t]
\centering
\includegraphics[scale=0.16]{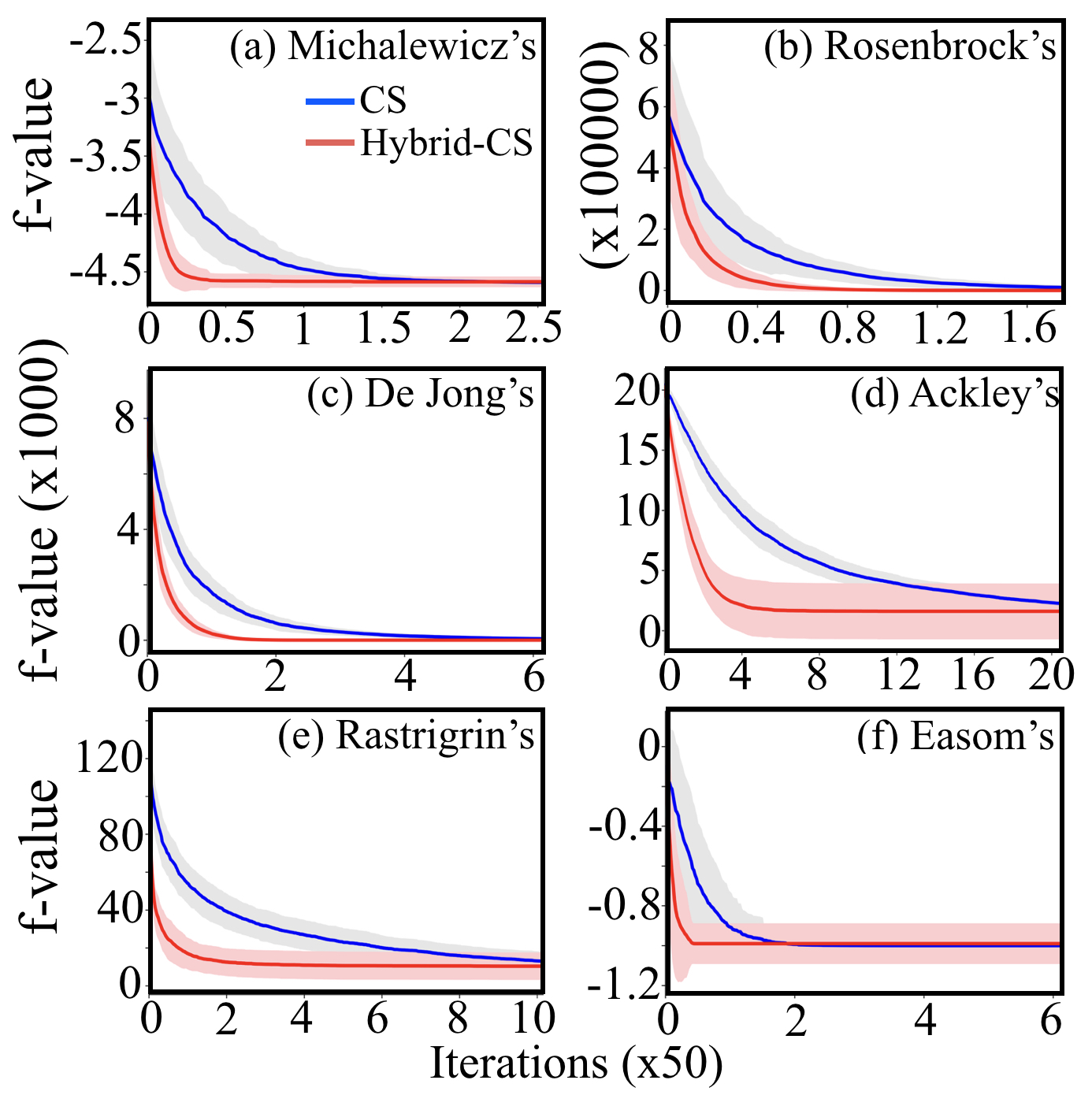}  
\caption{Hybrid CS and CS function values vs. iterations (objective evaluations) to reach optimum for six functions (defined in Supplemental Notes).  Each iteration is averaged over 100 runs. Error ($\pm1$ std) is denoted by shaded width of line.}
\label{fig:testfct}
\end{figure}

{\par}To show efficacy,  we applied hybrid CS and CS to 1D benchmark functions used in applied math (defined in Supplemental Notes). Simulations (mean over 100 runs) are shown for both algorithms versus iterations to reach optimum (Fig.~\ref{fig:testfct}). The algorithms converge to optimal values but at different rates. Hybrid CS outperforms CS in all cases, reducing evaluations by factors of $1.75$ (Michalewicz) to $8$ (Rastrigrin). MC is clearly a more efficient search of local minima than L$\'{e}$vy flights alone.\cite{Gutowski2001}

\begin{figure}[t]
\centering
\includegraphics[scale=0.16]{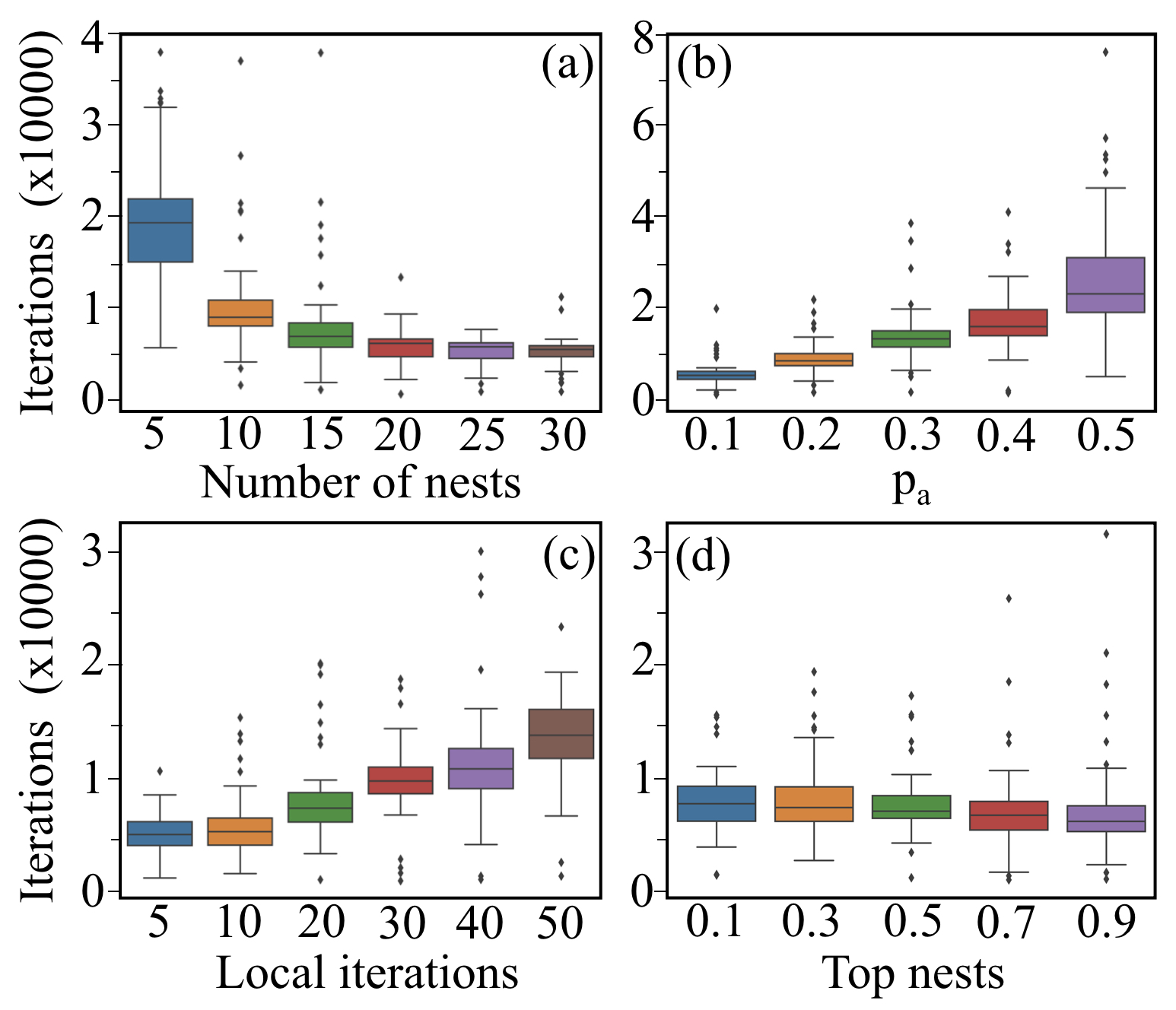}
\caption{Hybrid-CS performance: iterations vs. (a) \#nests, (b)  $p_a$, (c) \#MCiters, and (d) \#Top-Nests.}
\label{fig:HCS-tuning}
\end{figure}

{\par} The CS has two parameters: (1) \emph{number of nests} ($n$) and  (2) \emph{discard probability} ($p_a$).  Hybrid CS has two more: (3) \emph{fraction of top nests}  (\#Top-Nests) chosen for a MC step, and (4)  \emph{number of MC steps} (\#MCiters). 
Iterations to convergence versus parameter values are tested in Fig.~\ref{fig:HCS-tuning}: iterations are (a) roughly constant after $n$=$20{-}30$ and (b) increase linearly with $p_a$; while (c) \#MCiters increase roughly linearly after $10{-}20$; and (d)  \#Top-Nests passed in MC, with the rest untouched, has little effect on iterations. So, for least iterations to optimum, these results suggest:  $n\ge15$,  \#MCiters $\le n$, and  $0.1 < p_a <  0.4$. Parameter  were fixed for tests in Fig.~\ref{fig:testfct}: $n$=$15$, $p_a$=$0.25$, \#MCiters=15, and \#Top-Nests=0.3.

{\par}For any function with appropriate fitness, the hybrid CS out performs CS (which already bests all other evolutionary algorithms).  So, we employ hybrid CS for materials design to optimize SCRAPs using parameters found above. To download, see Code Availability.

\subsection{Solution Size and Fitness for SCRAPs}
\label{mat-sci-appl}
To assess hybrid-CS (pseudo)optimal SCRAPs, as in Fig.~\ref{fig:SCRAP5}, we need a fitness and size estimate of solution spaces in terms of $S$ sites and $N$ elements. We illustrated this for bcc  with $C$ cells built from 2-atom cubes such that $S=2C^3$; with $C=N$, SRO parameters can be exactly zero (homogeneously random) in a smallest-cell solution.  For a ternary (ABC),  quaternary (ABCD), and quinary (ABCDE), $S$ is a $54$-, $128$-, and $250$-atom cell, respectively. In terms of combinatorial coefficient $\prescript{S\mkern-0.5mu}{}{\cal C}_{{S/N\mkern+0.5mu}}$, the estimated configurations for site occupations are
\begin{itemize}{\small
    \item $ABC:   \prescript{54\mkern-0.5mu}{}{\cal C}_{18}\times \prescript{36\mkern-0.5mu}{}{\cal C}_{18} \approx 10^{23}$
    \item $ABCD: \prescript{128\mkern-0.5mu}{}{\cal C}_{32}\times \prescript{96\mkern-0.5mu}{}{\cal C}_{32}\times \prescript{64\mkern-0.5mu}{}{\cal C}_{32}\approx 10^{73}$
    \item $ABCDE:   \prescript{250\mkern-0.5mu}{}{\cal C}_{50}\times \prescript{200\mkern-0.5mu}{}{\cal C}_{50}\times \prescript{150\mkern-0.5mu}{}{\cal C}_{50}\times\prescript{100\mkern-0.5mu}{}C_{50}\approx 10^{169}$
     \item $ABCDE:  \prescript{500\mkern-0.5mu}{}{\cal C}_{100}\times \prescript{400\mkern-0.5mu}{}{\cal C}_{100}\times \prescript{300\mkern-0.5mu}{}{\cal C}_{100}\times\prescript{200\mkern-0.5mu}{}{\cal C}_{100} \approx 10^{415}$ }
\end{itemize}
A cell with bigger $S$ at fixed $N$ alters compositions in discrete but more refined ways, as evinced by the two $N=5$ sizes above; but times to render an optimal cell and use it is more challenging.  The solution space increase as the number of pairs {\small $\frac{1}{2}N(N-1)$} grow, and are needed for each atom and its pairs over range $R$ ($1$ to $5$ shells). 

{\par}  {\bf  Fitness --} An $N$-component CSA is characterized uniquely by $N-1$ one-site (occupation) probabilities $p_{\alpha}^i$ for species $\alpha$ and by $\frac{1}{2}N(N-1)$ two-site (pair) probabilities per neighbor shell, with definition (and sum rule):
\begin{equation} \label{EqP1}
  {\hat p}_{\alpha} = c_{\alpha} = \frac{1}{S} \sum_{i=1}^{S} p_{\alpha}^i  \;\;\;\;  \left(  \sum_{\alpha=1}^{N} {c}_{\alpha} = 1 \right)  ,
\end{equation}
\begin{equation}  \label{EqP2}
p_{{\alpha\beta}}^{ij} = p_{\alpha}^{i} p_{\beta}^{j}[1 - \alpha_{\alpha\beta}^{ij}]      \;\;\;\;    \left(  \sum_{\beta=1}^{N} {p}^{ij}_{\alpha\beta} =  {p}_{\alpha}^{i} \right)   .
\end{equation}
Here,  average compositions  (${\hat p}_{\alpha}=c_{\alpha}$) are given by sum over all sites S (with all species conserved).
The SRO parameters,  $\alpha_{\alpha\beta}^{ij}$, dictate pair probabilities $p_{{\alpha\beta}}^{ij}$ for having $\alpha$ ($\beta$) atom at site $i$ ($j$), and their values are bounded:\cite{Encyl2012} 
\begin{equation}\label{eq-SRO}
   -\frac{\text{min}(p_{\alpha},p_{\beta})^2}{p_{\alpha}  p_{\beta} }  \le \alpha_{\alpha\beta}^{ij}  \le +1  ,
\end{equation}
where $\alpha < 0$ indicates ordering-type SRO  (increased pair probabilities), whereas $\alpha > 0$ indicates clustering of like pairs  (decreased pair probabilities).  The final SRO for all sites and pairs qualify a model. 
So, a SCRAPs must be optimized with constraints for target SRO values, and, to avoid stagnation of solutions and senseless iterations (wasted computing), we place ``stop'' conditions on MC searches when SRO falls below discrete bounds set by the discrete cell $N$ and $S$ (see Methods).  Discrete  limits on floor/ceiling SRO values are exemplified for a non-cubic cell for a bcc equiatomic quinary in Supplemental Notes.

\subsection{Hybrid-CS  vs. MC-only Models}
\begin{figure}[h]
\centering
\includegraphics[scale=0.15]{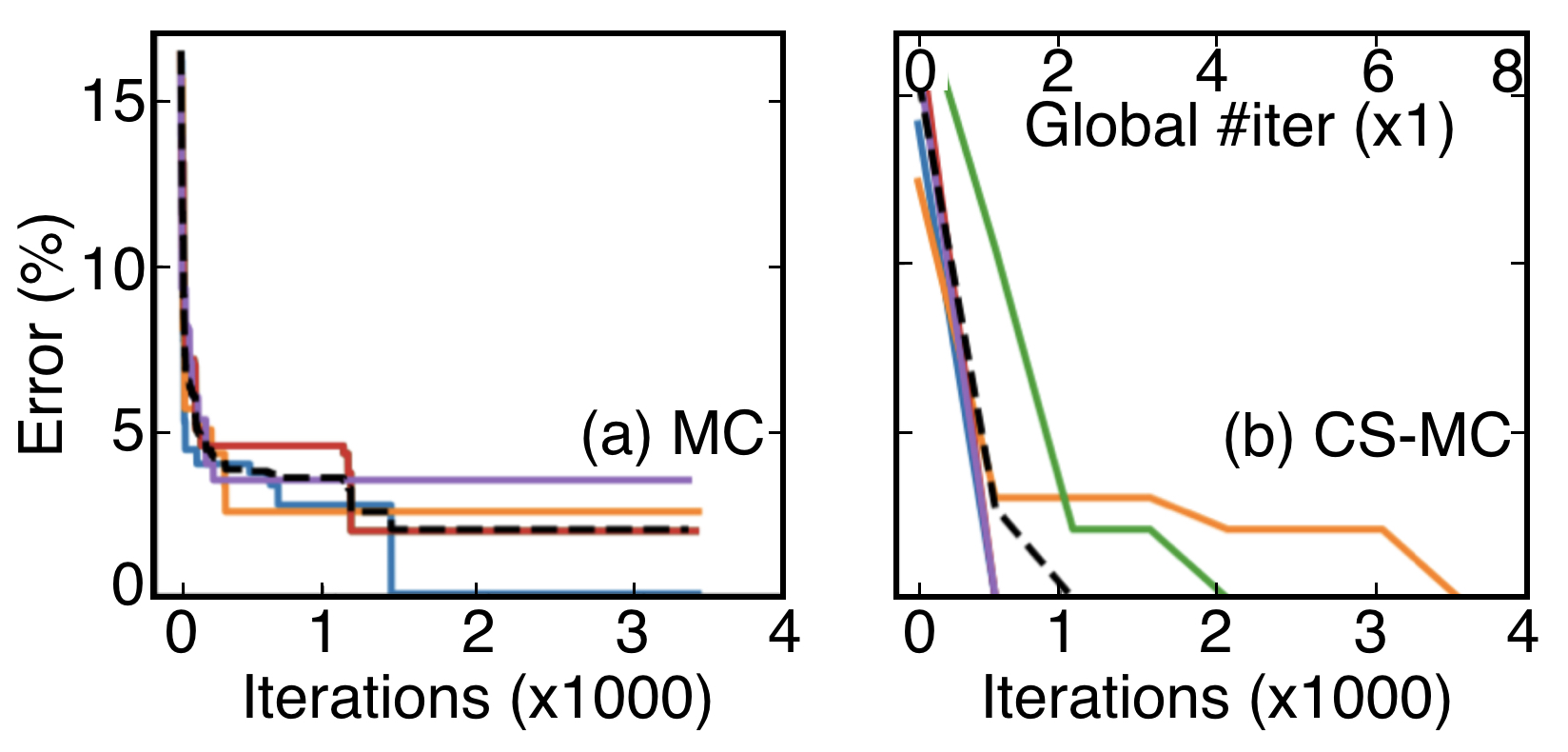}
\caption{For 54-atom bcc SCRAP, hybrid-CS vs MC-only optimization.
Runs (averages) are denoted by (dashed) lines.}
\label{fig:HCSvsMC}
\end{figure}

{\par} With MC stagnation addressed, hybrid CS enables {``on-the-fly''} generation of optimal SCRAPs to model CSAs with arbitrary concentrations, structures, and targeted atomic distributions.  For ease of plotting, we first use a ternary ($N$=$3$) with sites $S$=$54$ (no SRO) to compare hybrid CS vs. MC-only (Fig.~\ref{fig:HCSvsMC}). But, cells up to 2000 sites and 10 elements are timed in Table ~\ref{tab2}.  Background on MC-only generated cells is given in Methods.

{\par}Hybrid CS (0.3 mins) vs. MC-only (1440 mins or 1 day) from the ATAT code\cite{Walle2013} shows an impressive difference in timing (Table~\ref{tab2}), increasing significantly with larger $S$ and $N$. Hybrid CS was successful in every attempt (Fig.~\ref{fig:HCSvsMC}) to find the global (pseudo)optimum  -- zero SRO  for all pairs over 3 shells for every site -- irrespective of the initialization, albeit iteration count varied. 
MC-only failed to reach an optimum from stagnation in all cases but one (a random event). 
For hybrid-CS in larger cells with $N$= 3, 4, and 5, the SCRAPs have the targeted SRO and distributions are Gaussian  (see Supplemental Notes) -- a general result for all system sizes.

\begin{table*}[t]
\caption{Hybrid-CS timings ({\it in minutes}) to generate optimal cells using 24 nests, and solved with 1, 12, or 24 processors to show {strong scaling} in solution time for parallel cases. CS parameters were set to 10 iterations [converged in 3-5] and each iteration had up to 1,000  global and 750 MC iterations. For simplicity, cell sizes were set as $S=A \cdot N^3 $ [$A=2 (4)$ for bcc (fcc)] so SRO can be exactly zero. Although shells can be included to a range that avoids correlation from periodic boundaries, SRO was optimized over 3 shells (`*' denotes only 2 shells permitted). Comparisons are made with popular MC-based {\tt ATAT code}.\cite{Walle2013} Calculations used a cluster ({\tt Intel dual Xeon Gold 6130 at 2.1GHz/16-cores}). 
 }
\label{tab2}
\begin{tabular}{l|c|c|c|c|c|c|c}
\hline
Type  &  Species   &  $\#$ Atoms   & Hybrid-CS   & Hybrid-CS  & Hybrid-CS   & MC-only &   factor over \\
          &        (N)    &       (S)    & Serial  Timings  & Parallel Timings  & Scaling    &  Timings\footnote{Due to excessive computational demands, multinary results cannot be provided.}    & MC-only  \\
          &                 &                &   1 processor     & [12]  24 proc.      & [12] 24 proc. & via ATAT  &   [1]  24 proc. \\
 \hline
bcc  & 2    & 16       & 0.06    &  [0.008]  0.006          & [7.5]\, 10.0        & 0.4                    &   \\
bcc  & 2    & 32*     &  0.09    &  [0.010]  0.008         & [9.0]\, 11.3        &  0.5                   &   \\
bcc  & 2    & 40*     &  0.11    &  [0.013]  0.008         & [8.5]\, 13.8        & 105.2                &  [956]  13,150  \\
bcc  & 2    & 54*     &  3.16    &  [0.47]    0.21           & [6.7]\, 15.0        & --                      &    \\
bcc  & 3    & 54*      &   5.50   & [0.71]  {\bf 0.30}       & [7.7]\, 18.3        & $>$1,440 \, \, \, & [262] \, 4,800 \\
bcc  & 4    & 128    &  23.59  & [1.75]   {\bf 0.79}      & [13.5] 29.9  &    $>$10,000\footnote{For MC-only ATAT with higher-order cases, a sense of time is garnered from ternary 54-atom case.}{$^,$}\footnote{For MC-only ATAT 128-atom case, 7+ days but distributions were not assessed (MC often stagnates to non-Gaussian state).}  & [424] 12,658  \\
bcc  & 5    & 250    & 49.26  \,  & [3.57]  {\bf 1.60}           & [13.8] 30.8    &   unknown      &    \\
bcc  & 6    & 432    & 87.82  \,  & [6.56] {\bf 2.91}            &  30.2             &   --                 &    \\
bcc  & 7    & 686    & 143.50    & [10.62]  {\bf 4.73}         &  30.3             &   --                 &    \\
bcc  & 8    &1024   & 222.21    & [16.52]  {\bf 7.38}         &  30.1             &   --                 &    \\
bcc  & 9    &1468   & 319.67    & [23.94]  {\bf 10.82}  \,   &  29.5             &   --                 &    \\
bcc  &10   &2000   & 446.53    & [33.74] {\bf 15.48}  \,    &  28.8             &   --                 &    \\
\hline
fcc   &  2  & 32       & 0.12      & [0.012]  0.008         & [10.0] 15.0           & --                  & --     \\
fcc   &  2  & 108     & 18.15    & \,[1.38]  {\bf 0.62}    & [13.4] 29.3          & --                  & --    \\
fcc   &  3  & 108     & 26.24    & \,[1.99]  {\bf 0.86}    & [13.2] 30.5          & --                  & --    \\
fcc   &  4  & 108     & 31.12    & \,[2.30]  {\bf 1.01}    & [13.5] 29.8          & --                  & --     \\
fcc   &  4  & 256     & 70.49    & \,[5.28]  {\bf 2.30}    & [13.4] 30.6          & --                  & --     \\
fcc   &  5  & 500     & 149.71  & [10.99]  {\bf 4.89}\,  & [13.6] 30.6          & --                  & --     \\
 \hline
AXB$_{3}$\footnote{Cubes of A (organic cation) at body-center, X (mixed inorganic cation, Pb or Sn) at corner, and B (halide) at face-centers.}   
          & 3     &  10      & 1.44 & [0.29] 0.15 &  [4.9] 9.6 &  -- & -- \\
   \hline
\end{tabular}
\end{table*}
%

\subsection{Hybrid-CS SCRAPs: Timings \& Scaling}
{\par}Timings for hybrid-CS-created SCRAPs (Table~\ref{tab2}) show markedly reduced times compared to MC-only, which suffer stagnation as $S$ and $N$ increase. For a binary 40-atom cell, MC-only needed $105$ mins while hybrid CS in serial mode used $0.11$ mins ($<0.01$ mins in parallel).

{\par}Hybrid-CS timings exhibit strong scaling versus number of processors ($n_{proc} \le n_{nests}$); for 24 nests and 24 processors, we observe a reduction factor  of 30  (Table ~\ref{tab2}). Scaling is detailed ($t$ vs $n_{proc}$) in Supplemental Notes, with >200\% decrease for each doubling of processors. Parameters were set to 10 optimization steps (typically converged in 3-5) with a solution for each step having up to 1,000  Levy and 750 MC searches (iterations) at each step.

{\par} As SRO qualify solutions, we note that all multinary SCRAPs in Table ~\ref{tab2} have specified value of zero for 3 shells about all sites (worst error $<10^{-3}$ for two pairs in 3$^{rd}$ shell in 10-element, 2000-site case).  Thus, hybrid CS yields optimal models in minutes.

{\par} The scaling of times is better than anticipated from our estimated sizes due to a limit on range on SRO. Typically the range is limited in a solid solution, except near a phase transition where SRO diverges (or if Fermi-surface nesting operates\cite{Singh2018b}).
Optimization at each site over a few shells ($Z_{tot}$ atoms) is then usually valid [to our benefit] and should scale as cell size $S$. For timings, we chose $S=A \cdot N^3 $ [where $A=2 (4)$ for bcc (fcc)].   The relative time $t_{rel}$ for different symmetries is then easy to assess: For fixed $N$ and $R$ (range of SRO, say 3 shells), we find $t^{N}_{rel} = \frac{A_{fcc}}{A_{bcc}} \cdot \frac{Z^{fcc}_{tot}} {Z^{bcc}_{tot}}=\frac{4}{2} \cdot \frac{42}{26}=3.2$; it is $3$ in Table~\ref{tab2}. Timings for fixed symmetry, say two sizes of bcc cells, should scale as  $t^{sym}_{rel} = \frac{S_{2}}{S_{1}} \cdot [\frac{N_{2}} {N_{1}}]^{1/4}$, as found in Table~\ref{tab2}.  So, the algorithm scales linearly with system size $S$.

{\par}Thus, hybrid-CS SCRAPs to address concentration-dependent CSA properties are obtained rapidly. Six bcc 250-atom quinary SCRAPs, like $A_{x}(BCDE)_{1-x}$ vs. $x$ composition, for example, are found in minutes.  And,  any SRO values may be targeted, as SRO in alloys can lower the enthalpy or drive elemental surface-enrichment. Smaller $S$-atom cells with larger $N$ can be obtained, but zero SRO will not be possible at all compositions. 

{\par}Clearly, then, for an arbitrary multinary with any  sample size (SCRAP) and for any targeted disorder (SRO), optimization runtime for parallel-mode hybrid CS is superior or the only alternative for design of CSAs. We have eliminated model generation as a bottleneck, which had larger times than DFT calculations.

\subsection{Example Real Alloy Applications}\label{SCRAPS-APPS}
We constructed SCRAPs to assess formation energy (E$_{form}$) versus SRO parameters (observed or trial $\alpha_{\mu\nu}^{n}$).  
We assess relative energy ($E$) versus lattice constants ($a$) and equilibrium values  (${\bar a}$), along with atomic displacement $\{u_{i}\}$ distributions for binaries to quinaries. 
We employed an all-electron KKR Green's function method\cite{JohnsonCPA1,Alam2012} and pseudopotential {\emph Vienna Ab-initio Simulation Package} (VASP) \cite{VASP1,VASP2} to calculate E$_{form}$ vs SRO and ${\bar a}$, compared to experiments.  See Methods Section for details and Supplemental Notes for supporting results.

\begin{figure}[!h]
\centering
\includegraphics[scale=0.18]{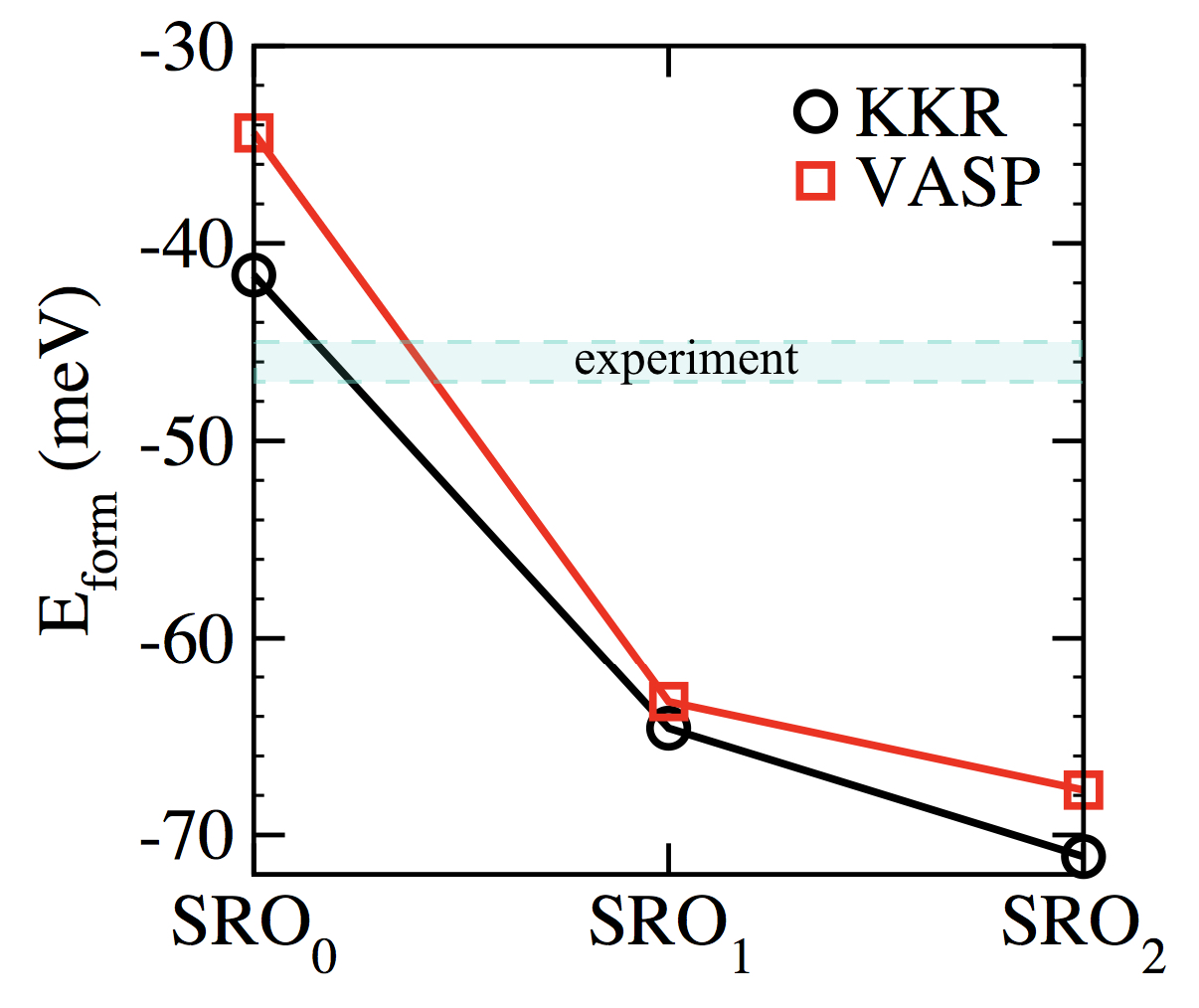}
\caption{E$_{form}$ vs. SRO parameters for Cu$_{0.75}$Au$_{0.25}$ in $108$-site SCRAP from KKR (circles) and VASP (squares). 
Measured $\alpha$'s (3 shells)\cite{Moss1964}  at 
$495^o$C SRO$_0$=$\{0.0, 0.0, 0.0\}$;
$450^o$C SRO$_1$=$\{-0.195, +0.215, +0.003\}$; and  
$405^o$C SRO$_2$=$\{-0.218, +0.286, -0.012\}$.  
For SRO$_0$, a measured range\cite{Cu3Au-1} of E$_{form}$ is shown (dashed-horizontal band).   }
\label{fig:Cu3Au-SRO}
\end{figure}

{\par} {\bf CSAs with SRO:} As hybrid CS works for any $\frac{1}{2}N(N-1)$ SRO pairs, we exemplify fcc Cu$_3$Au ($N$=2) for ease of presentation (one CuAu SRO value per shell, $\alpha^{n}$) and there is experimental data. SCRAPs with specified SRO (optimized in $0.6$~mins, Table~\ref{tab2}) are used to mimic (a) a homogeneously random state at $495^o$C ($\alpha^{n}$=0) and alloys with observed $\alpha^{n}$ values \cite{Moss1964} at (b) $450^o$ and (c) $405^o$C.  In Fig.~\ref{fig:Cu3Au-SRO}, E$_{form}$ vs. SRO is plotted.   As found for any $N$ and $S$, the atomic displacements from ideal sites have zero mean (see Supplemental Notes).

{\par} KKR and experiment with no SRO agree well ($3$-$5$~meV difference). 
Both methods show similar trends, but KKR includes known alloying core-level shifts, explaining the VASP higher value.
 A 30 meV/atom gain is found with SRO (lower entropy).
KKR ${\bar a}$ without SRO is $3.765 {\AA}$, agreeing with the $3.749\AA$ observed, \cite{Flinn1960,Cu3Au-1,Cu3Au-2} a small $0.43{\%}$  mismatch.
A disordered alloy with SRO is $3.755 {\AA}$, closer to the ordered alloy $3.743\AA$.
VASP ${\bar a}$ with no SRO is $3.823 {\AA}$ (with SRO is $3.816 {\AA}$) for a $2{\%}$  mismatch.

\begin{figure*}[!t]
\includegraphics[scale=0.22]{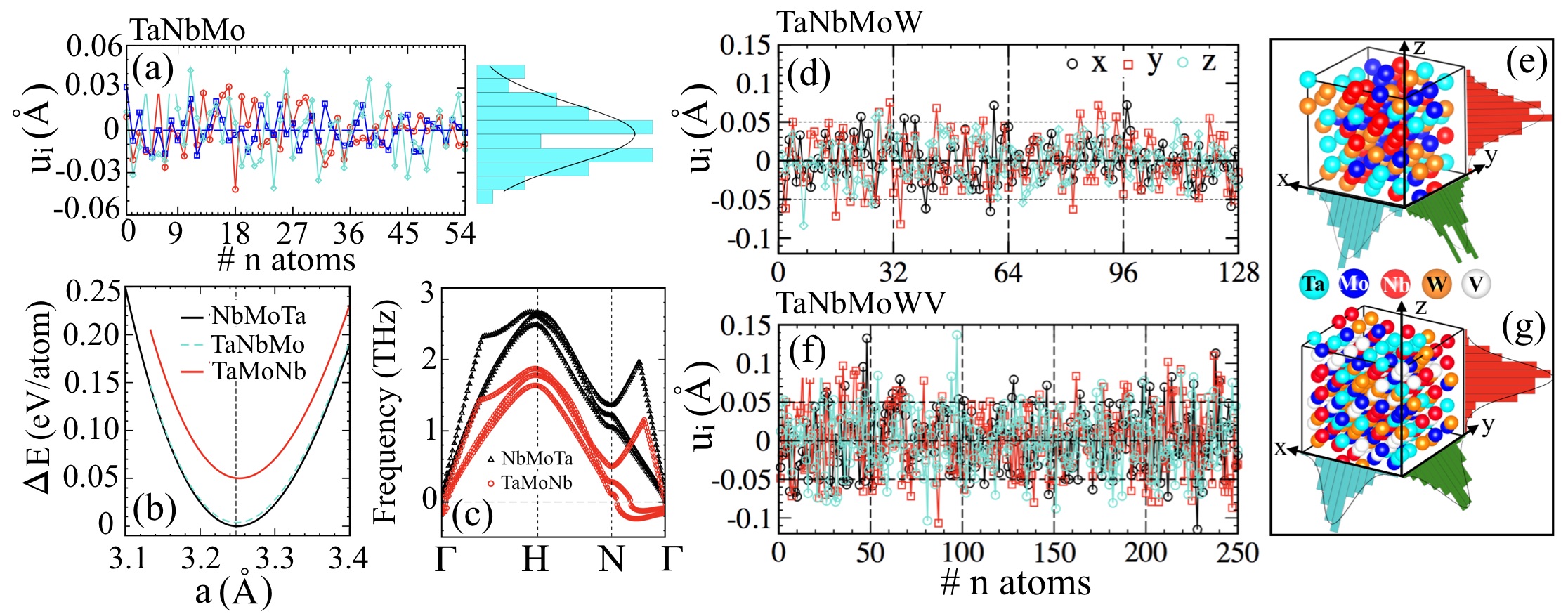} 
\caption{SCRAPs for refractory (a-c) TaNbMo (54-atom), (d-e) TaNbMoW (128-atom), and (e-f) TaNbMoWV (250-atom).
(a), (d) and (f) Components $u_{i}$  of vector atomic displacements (mean $0$) relative to their average (scattering) lattice position.
For clarity, only $u_z$ distribution is shown in (a), and the average squared-displacement (mean of $0.027\AA^2$) is shown in Fig.~S4. 
(b) Energy for TaNbMo relative to lowest energy for the 3  A, B, C assignments, and, as required, $\bar{a}=3.248\AA$ is the same for each. 
(c) Acoustic phonons for stable NbMoTa (positive-definite frequencies) and unstable TaMoNb (imaginary frequencies, plotted as negative).
Distributions  for $u_{i}$ for (e) TaNbMoW with $\bar{a}=3.247\AA$ and (g) TaNbMoWV with $\bar{a}=3.198\AA$.    }
\label{fig:TaNbMo}
\end{figure*}

{\par}{\bf Distributions \& Averages:}  To simplify presentation, we assess VASP $E$ vs. $a$ and displacements $\{u_{i}\}$ for NbMoTa 54-atom SCRAP (Fig.~\ref{fig:TaNbMo}a-c).  
At $\bar{a}$,  energy is $-64.4$~meV/atom for volume-relaxed (ideal sites), and $-80.5$~meV/atom with atom-relaxed ($-16$~meV gain).
Vector  $\{u_{{x,y,z}}\}$ sum to zero individually and are Gaussian distributed, as required by CSA symmetry, giving $\bar{a}$ as the diffraction value. Mean-squared displacements  determine the Debye-Waller factor (Supplemental Notes) that describes attenuation of x-ray, neutron, or electron scattering caused by thermal motion giving background diffuse intensity from inelastic scattering.  Diffraction on ``large'' samples (e.g., 1 cm$^3$) obtain ``self-averaged'' properties, as Avogadro's number of local configurations are simultaneously sampled. 
Similar results are found for any $N$ and $S$ (Figs.~\ref{fig:TaNbMo}d-g).  For quaternary TaNbMoW, volume- ($-63.3$~meV/atom) and fully-relaxed ($-74.5$~meV/atom) gives $-11$~meV gain.  For quinary TaNbMoWV, volume- ($-105.5$~meV/atom) and fully-relaxed ($-126.3$~meV/atom) both show a significant stability gain with vanadium addition, and larger ($-21$~meV) gain from displacements. Displacements clearly increase with complexity but more with V alloying (Figs.~\ref{fig:TaNbMo}d,g), enhancing stability, lattice distortions and mechanical behavior, as discussed below.

{\par} {\bf Configurations:} SCRAPs provide  good averages if a cell is large enough (``infinite'' is exact, but impractical); otherwise configurational averaging may be warranted. In principle, all configurations should be sampled:  good, bad, and ugly (leading to an average ideal lattice); not just a relaxed, low-energy one, as often chosen in literature. 
SCRAPs (before relaxations) are for arbitrary choices of atomic site occupations, as it is just one representative configuration out of many, so a model must be qualified. To complete a model, atom types Nb, Mo, or Ta must be assigned to A,B, or C sites. For example, formation energy after relaxations may be fine but phonons may exhibit lattice instabilities (indicated by ``imaginary'' phonon frequencies).  A statistical average governs Nature's reality, and an instability is controlled by environments around each atom. So, to eliminate the instability, you must use a larger SCRAP to improve the statistical ``self-average'', or you may be able to simply swap atom types in a given small SCRAP to eliminate a local instability.
For instance, {if we assign Nb, Mo, and Ta to A, B, and C sites, respectively, we find a minimum energy and stable lattice, i.e., positive phonon frequencies (Fig.~\ref{fig:TaNbMo}c). From these results we can assess the alloy properties, e.g., $\bar{a}$ is $3.248\AA$. Yet, with $A \leftrightarrow C$ (i.e., Nb $\leftrightarrow$ Ta), we find a higher ($+0.05$~eV) energy  and unstable phonons (Fig.~\ref{fig:TaNbMo}c), suggesting that this model is in general too small, and care should be taken.}

{\par} {\bf Lattice Distortions:} Each atom in a CSA has a different chemical environment that can cause lattice distortion, e.g., from atomic size differences.\cite{Yeh2013} However, the effect of lattice distortion on CSA mechanical response remains less explored due to absence of computationally efficient models. In SCRAPs the lattice distortion in refractory CSAs can be tuned by changing the local environment to enhance mechanical response (intimated in Fig.~\ref{fig:TaNbMo}a,d,f), an effect observed in ultra-strong ternaries.\cite{Sohn2019} 
Rather than size difference, embodied to zeroth-order in a solid-solution's electronic bandwidths [the electronic origin of Hume-Rothery's size-effect rule\cite{Pinski1991}], strength enhancement correlates with electronegativity difference between elements (on Allen scale for solids, with V largest), where largest bond distortions occur around V sites (Supplemental Notes).

%
\section{Discussion}
We presented a hybrid Cuckoo Search (CS) -- a modified evolutionary algorithm -- utilizing L$\'{e}$vy flights for global exploration coupled with Monte Carlo for local searches, with no stagnation. 
To accelerate materials design, our hybrid CS overcomes large, discrete combinatorial optimization by ultrafast global solutions for extremely large solution spaces, scaling linearly in size and strongly in parallel. 
The hybrid CS was used to generate SCRAPs to model complex multinary solid-solution alloys with targeted degree of disorder -- a type of NP-hard combinatorial problem.  We assessed stability and properties for solid solutions and discussed qualification of the models.  For smaller design spaces, optimal SCRAPs are created four orders of magnitude faster than existing methods.  For larger cases, no comparison to current methods were possible, but only minutes are needed up to 2000 atoms and 10 elements with specified SRO (Table~\ref{tab2}).  

{\par} Having saved orders of magnitude in model generation, DFT times are again a major issue. But, a savings in DFT time is also possible.  As displacements $\{u_{i}\}$ must have zero mean in any disordered alloy,  the equilibrium (average) volume must be mathematically identical to that of the ``ideal'' (diffraction) lattice (e.g., Fig.~\ref{fig:TaNbMo}a).  As such, relaxations for any sized SCRAP with any SRO need only be performed at the equilibrium volume  to assess properties and trends (e.g., Fig.~\ref{fig:TaNbMo}b). As DFT methods typically scale as $S^3$, a huge savings is realized.

{\par}For arbitrary multinary disordered alloys, hybrid CS generates accurate ``on-the-fly'' models (SCRAPs) in an exponential solution spaces in a few minutes, truly enabling rapid design of properties and trends, including for surfaces, catalysis, and oxidation.   With this design bottleneck removed, computational alloy design can be performed that is currently impossible or impractical.  A hybrid CS should also improve  optimization applications found in other fields, such as manufacturing, commerce, finance, science, and engineering. The hybrid CS code used to get bcc, fcc, or hcp SCRAPs and timings in Table~\ref{tab2} can be download (see Data Availability).

\section*{acknowledgements}
R.S. was supported in part by D.D.J.'s F. Wendell Miller Professorship at ISU. Work at Ames Laboratory was funded by the U.S. Department of Energy (DOE),  Office of Science, Basic Energy Sciences, Materials Science \& Engineering Division. Ames Laboratory is operated for the U.S. DOE by Iowa State University under Contract DE-AC02-07CH11358. G.B. was funded by the National Science Foundation through award 1944040.

{\noindent} 
\section*{Author Contributions}
DDJ proposed and supervised the project. RS wrote the SCRAPs generation code using hybrid-CS algorithm. RS and AS did initial testing. DDJ developed strong-scaling algorithm and scaling analysis. PS and RS implemented SCRAPs optimization with parallelization and catalogued timings. PS completed DFT and phonon calculations, and analyzed with DDJ. PS got code running on CodeOcean. After RS, AS, PS, and GB drafted initial manuscript, DDJ prepared the final manuscript with approval from all the authors.

{\noindent}
\section*{Corresponding Author}
D.D.J. (ddj@iastate.edu; ddj@ameslab.gov).


\section*{Methods}
\subsection*{Cuckoo Search}
{\par} The standard CS  is based on the brood parasitism of a female Cuckoo bird that specialize in mimicking the color and pattern of a few host species -- using three idealized rules: (i) a cuckoo lays an egg in a randomly selected nest; (ii) the nest with highest-quality egg (fitness) survives and is forwarded to the next generation; (iii) the host bird can discover the cuckoo egg with a probability p$_{a}$ $\in$ (0,1), and, once discovered, it dumps either the nest or the cuckoo egg. The advantages are: (a) global convergence has higher success relative to other approaches, (b) it employs local and global searches controlled by a switching parameter, and (c) it uses L$\'{e}$vy flights to scan solution space efficiently -- no random walks, so better than Gaussian process.\cite{Yang2009,Yang2013} 

\subsection*{Hybrid Cuckoo Search}\label{hybridCS}
Our hybrid-CS schema reaps benefits of traditional MC for local optimization alongside the CS schema for global optimization utilizing multiple nest explorations via L$\'{e}$vy flight. A global CS removes a fraction of nests, $p_a$, with worst fitness (a nest represents a lattice configuration); and it signifies the probability of finding an alien nest.\cite{Yang2009}
As shown in the pseudo-code, we replace the local search in CS Algorithm~\ref{Algorithm1} by MC and create a hybrid CS given in Algorithm~\ref{Algorithm2} below, where the global search uses multiple-nest explorations.  Notably, to improve the hybrid CS, we tested various MC approaches, even simulated annealing, and the basic version embodied in Algorithm~\ref{Algorithm2} was superior to all others for local optimization. The basic MC version (shown between $\ll$MC$\gg$ symbols in Algorithm~\ref{Algorithm2})  is: (1) Obtain a nest from the sample of nests. (2) Randomly swap a pair of lattice occupations. (3) If Fitness\textsubscript{new} < Fitness\textsubscript{old}, the Accept swap, or (4) Else Reject; Switch; and Continue (Go to step 1).

{\par}``Local'' MC iteration chooses a fraction of nests to optimize based on a nest's value of fitness and a fraction equal to {\em{top nests}} $\in \{0, 1\}$. Aside from ``{top nests}'', the local MC depends on $mc_1 \in \{0, 1\}$ and $mc_2 \in \{0, 1\}$ that are used to optimize the value of step size, $\delta x = \sigma*randn$ by altering the value of $\sigma$ ($randn$ is a value from a standard normal distribution). For local optimization, number of acceptances/rejections are counted and, depending on their value, the value of $\delta x$ alters. The other parameters are $a (b) > 1$, the increase/decrease increment in $\sigma$. By collecting the number of acceptances/rejections, we increase/decrease $|\delta x|$ to get local optimized value faster.

\begin{algorithm}[H]
    \SetAlgoLined
    \SetNoFillComment
    \DontPrintSemicolon
    \SetKwInOut{Input}{Input}
    \SetKwInOut{Output}{Output}
    \Input{Fix input \& identify optimization variables}
    \Output{Optimized solution}
    Initialize nests \\
    \While  {iteration $<$ Global maximum number}{
        Create new nests using L$\'{e}$vy Flight ({Global Search})\\
        Calculate fitness $F$ of the nests\\
	Choose a nest randomly \\
        \If {$F_{old} < F_{new}$}{replace $nest$ with the new cuckoo}{
        Discard fraction $p_{a}$ of worst nests \& built new ones \\
        Keep best nests with the best results \\
        Rank the solutions \& find the current best}}
	{Return the best solutions}
    \caption{Cuckoo Search Algorithm}
    \label{Algorithm1}  
\end{algorithm} 

\begin{algorithm}[H]
 \SetAlgoLined
    \SetNoFillComment
    \DontPrintSemicolon
    \SetKwInOut{Input}{Input}
    \SetKwInOut{Output}{Output}
    \Input{Fix input and optimization function}
    \Output{Optimized solution}
    Initialize nests \\
    \While{iteration $<$ Global maximum number}{
        Create new nests using L$\'{e}$vy Flight ({Global Search})\\
        Calculate fitness $F$ of the nests\\
        Choose fraction of nests with best fitness (\textbf{top nests})\\
     $\bf{\ll MC\gg}$  Search using Monte Carlo ({Local Search}) \\
          \ForEach {nests $ \in$ \textbf{top nests}}{
            acceptance = 0\\
            rejections = 0\\
        \While{iteration $<$ Local iterations } {           
            Calculate delta step ($\delta x = \sigma*randn$)\\
            Perturb nests with $x_{i} + \delta x$ \\
            Calculate fitness, $F(x_{i} + \delta x)$ \\
            Calculate $\delta F = F(x) - F(x_i + \delta x)$ \\
            \If{$\delta F > 0$}{Perform the switch\\ acceptance += 1}
                        \Else{rejections += 1}
            \If{ acceptance > $mc_1$* Local iterations}{
            {$\sigma$ = $\sigma*a$}}
            \If{ rejections > $mc_2$* Local iterations}{
            {$\sigma$ = $\sigma/b$} } }  }
    $\bf{\ll MC\gg}$  Discard fraction $p_{a}$ of worst nests  \\
    Rank the solutions \& find the current best}
	{Return the best solutions}
    \caption{Hybrid CS Algorithm}
    \label{Algorithm2}
\end{algorithm}

\subsection*{Bounded Discrete Searches: No Stagnation}
A SCRAPs must be optimized with constraints for target SRO values:  
\begin{equation}\label{eq-MIN}
\begin{aligned} 
& {\text{minimize}}
& & \mathrm \sum |{\hat{\alpha}}^s_{\alpha\beta} - d^s_{\alpha\beta}|\\
& \text{subject to}
& & \sum_{\alpha=1}^{N} {p^i_{\alpha}} = 1 \,\,\,   \text{and}  \sum_{\beta=1}^{N} {{p}_{ \alpha\beta}^{ij} = {p}^i_{\alpha}}
\end{aligned}
\end{equation}
${\hat{\alpha}}^s_{\alpha\beta}$ refers to average SRO for the $s^{th}$-shell for $(\alpha,\beta)$ pair. 
For {\small $\frac{1}{2}N(N-1)$} pairs, $d^s_{\alpha\beta}$ is the target SRO values.  
Final SRO for all sites and pairs qualify the model. 

{\par} To avoid stagnation of solutions we place ``stop'' conditions on MC (local) searches when SRO falls below the discrete bounds set by the cell $N$ and $S$.   Such criteria avoid senseless iterations (wasted computing), working well when combined with a CS that guarantees global (pseudo-optimal) convergence in a range $R$.  Discrete limits for SRO parameters from Eqs.~\ref{EqP1}--4 are: 
\begin{equation}
    (n_{\alpha} - \frac{\lfloor{g^{s}_{\alpha\beta}}\rfloor}{n_s c_{\beta}}) \frac{1}{n_{\alpha}}  \leq    \alpha_{\alpha\beta}^{s} \leq  (n_{\alpha} - \frac{\lceil g^{s}_{\alpha\beta} \rceil }{n_s c_j}) \frac{1}{n_{\alpha}}  ,
\end{equation}
with radial distribution function ($g^{s}_{\alpha\beta}$), number of atoms  in shell $s$ ($n_s$) and type $\alpha$ ($n_\alpha$). $\lfloor\text{ }\rfloor$ and $\lceil\text{ }\rceil$ represents the decimal at the lower (floor) and higher (ceiling) integer values, respectively. We use distance of SRO $\alpha_{\alpha\beta}^{s}$ from one of these values for a ``stop'' criteria, i.e.,
\begin{eqnarray}
 \sum |{\bar \alpha}_{\alpha \beta}^{s}- (n_{\alpha} - \frac{\lfloor{g_{\alpha \beta}}\rfloor}{n_s c_{\beta}})\frac{1}{n_{\alpha}}  | \leq \epsilon_1  \nonumber   \\  
 \sum |{\bar \alpha}_{\alpha \beta}^{s} - (n_{\alpha} - \frac{\lceil g_{\alpha \beta} \rceil }{n_s c_{\beta}}) \frac{1}{n_{\alpha}}   | \leq \epsilon_2
\end{eqnarray}
where $\epsilon_1$ and $\epsilon_2$ are predefined values.
Choosing $S$ and $N$ to set discrete $p_{\alpha}^i$ and specifying $\frac{1}{2}N(N-1)$ target values for  $p_{{\alpha\beta}}^{ij}$,  the final values of SRO for all sites and atom pairs qualify the model fitness. 
For the reader, discrete  limits on floor/ceiling SRO values are exemplified in a $3\times3\times5$ bcc supercell for an equiatomic quinary in the Supplemental Notes with values compared in Table~S2.

\subsection*{Density Functional Theory (DFT)}
{\par} For Binary Cu$_{3}$Au, VASP calculations were done on a $108$-atom SCRAPs with SRO.  Structures were relaxed by choosing 350 eV plane-wave energy cutoff, $3{\times3\times}3$ Monkhorst-Pack k-mesh\cite{HJM1976} for Brillouin-zone integrations and PBE exchange-correlation DFT functional.\cite{PEB1994} Total-energy calculations were done at denser ($7{\times7\times}7$) $k$-mesh. By definition, $E_{form}=E_{tot} - \sum_{i}n_{i}E_{i}$, where E$_{tot}$ [E$_{i}$] is the total energy of the alloy [pure elements `i'], n$_{i}$ number of sites per element in a supercell. For same k-meshes,  KKR\cite{JohnsonCPA1} was also used for $E_{form}$.  Self-consistent charge densities were obtained from the Green's function by complex-energy (Gauss-Legendre semicircular) contour integration with  24 energies in a spherical-harmonic basis, including $s$, $p$, $d$, and $f$ orbital symmetries.\cite{Alam2012}  Core eigenvalues were from Dirac solutions, and valence used scalar-relativistic solution (no spin-orbit coupling).

{\par}  {\it For Ternary to Quinary Alloys:} A 54-atom TaNbMo SCRAPs was relaxed in VASP using 350 eV plane-wave energy cutoff, $8{\times8\times}8$ $k$-mesh and PBE exchange-correlation.\cite{PEB1994} 
Energy and force convergence criteria for phonons were set very high to $10^{-7}$ eV and $10^{-6}$~{eV/\AA}, respectively. A finite-displacement  phonon method (set to $0.03\AA$)  was employed using PHONOPY.\cite{AT2015} Phonon dispersion was plotted along high-symmetry  Brillouin zone directions ($\Gamma$-H-N-$\Gamma$).  Compared to ternary, the only difference for 128-atom  TaNbMoW and 250-atom TaNbMoWV was the k-mesh, i.e., $5{\times5\times}5$ and $2{\times2\times}2$, respectively.

\subsection*{Related Cellular Techniques}
A supercell to mimic random alloys is not a new idea. Structural models are often constructed by specifically occupying sites of a finite-sized periodic cell.  For Metropolis MC methods,\cite{Ceguerra2012} including simulated annealing, potential energies serve as a fitness criterion for acceptance of a trial move, yet solutions for global optima often stagnate even in problems that are not large.\cite{Gutowski2001} 
We discuss fitness for SCRAPs in Results.

{\par}The original Special Quasi-random Structure (SQS) used Ising-like MC to find supercells that mimic zero atomic correlations in the alloy by arranging atoms in particular ordered layers depending on the number of sites and atom types;\cite{Zunger1990} and, in some cases, there were more than one configuration for a fixed number of sites, requiring an average. Such SQS did not have proper lattice symmetry (like bcc), and so atomic displacements could not sum to zero as required by symmetry, in contrast to SCRAPs.  Recently, the SQS algorithm was implemented using a stochastic MC approach\cite{Walle2013} to determine a sample configuration allowing a supercell with arbitrary number of base units, such as $N_1 \times N_2 \times N_3$  fcc 2-atom cells, as done for SCRAPs.  However, as $N$ or $S$ increase, MC-only times become impractical and solutions stagnate. Other implementations of MC-only approach with arbitrary number of base units was applied, although some results were correlated through use of improper boundary conditions.\cite{Song-Vitos2017} 

{\par} In principle, hybrid-CS  and MC-only schemas should get the same supercells for specific cases. However, our hybrid-CS algorithm completely avoids stagnated solutions  and the timings are significantly reduced in serial-mode, and exceptionally  reduced in parallel-mode  (see Results).  Moreover, we can optimize any sized SCRAP for any number of elements and for any targeted disorder (SRO), see Results.

{\noindent} 
\section*{Data Availability}
Supporting data for all figures are available,\cite{SCRAPs2020} see files {$SourceData\_Fig.xls$ for figures 1-6, with additional files for Supplemental Notes and other data also provided.

{\noindent} 
\section*{Code Availability}
Open-source codes are available for Hybrid-CS SCRAPs\cite{SCRAPs2020} and for Hybrid CS for 1D functions.\cite{HybridCS-test}

{\noindent} 
\section*{Competing interests}
The authors declare no competing interests.


\end{document}